\documentclass[10pt,a4paper,twocolumn]{article}
\usepackage[utf8]{inputenc}
\usepackage{amsmath}
\usepackage{amsfonts}
\usepackage{amssymb}
\usepackage{graphicx}
\usepackage{authblk}
\usepackage{geometry}
\usepackage{color}
\usepackage{float}
\usepackage{miller}

\title{Low Temperature Ageing Behaviour of U-Nb $\gamma^{o}$ Phase Alloys}
\author[1]{JE Sutcliffe}
\author[1]{CP Jones}
\author[1]{JE Darnbrough}
\author[1]{KR Hallam}
\author[1]{RS Springell}
\author[2]{P Ryan}
\author[2]{T Cartwright}
\author[1]{TB Scott}
\affil[1]{Interface Analysis Centre, HH Wills Physics Laboratory, University of Bristol, Tyndall Avenue, Bristol BS8 1TL, UK}
\affil[2]{AWE, Aldermaston, Reading, Berkshire RG7 4PR, UK}

\date{\today}
\begin{document}
 
\maketitle

\abstract{Ageing mechanisms of the U-7\%wtNb alloy have been investigated on samples exposed to temperatures of 150$^{o}$C for up to 5000\,hours. A variety of surface and bulk analytic techniques have been used to investigate microstructural, chemical and crystallographic changes. Characterisation of the alloy's evolving behaviour was carried out through secondary electron microscopy, energy dispersive x-ray spectroscopy, electron backscatter diffraction, transmission electron microscopy and x-ray diffraction. Vickers hardness testing showed evidence of a strong thermal hardening relationship with ageing. The mechanism responsible for these changes is thought to be a stress-induced isothermal martensitic transformation; a displacive reaction, in which correlated shuffling of atoms creates a phase change from $\gamma^{o}$ to $\alpha''$ without chemical species redistribution.}

\section{Introduction}
Considerable interest exists for uranium-based materials exhibiting a body centred structure. Besides the most common use of uranium alloys in defence applications, $\gamma$-phase uranium alloys have received huge interest as low-enriched, high-density fuels for use in research and test reactors \cite{Snelgrove1997,Meyer2002,Savchenko2010,Sinha2009,VandenBerghe2008a,Clarke2015}.

Effective, long-term storage and stockpiling of nuclear materials demands a thorough understanding of corrosion and ageing behaviour of uranium based materials. Alloying uranium with niobium has been shown to improve corrosion resistance, irradiation resistance, ductility and introduce shape memory effects \cite{Wheeler2009,Jackson1970,Manner1999}.

\begin{figure}
\centering
\includegraphics[width=\linewidth]{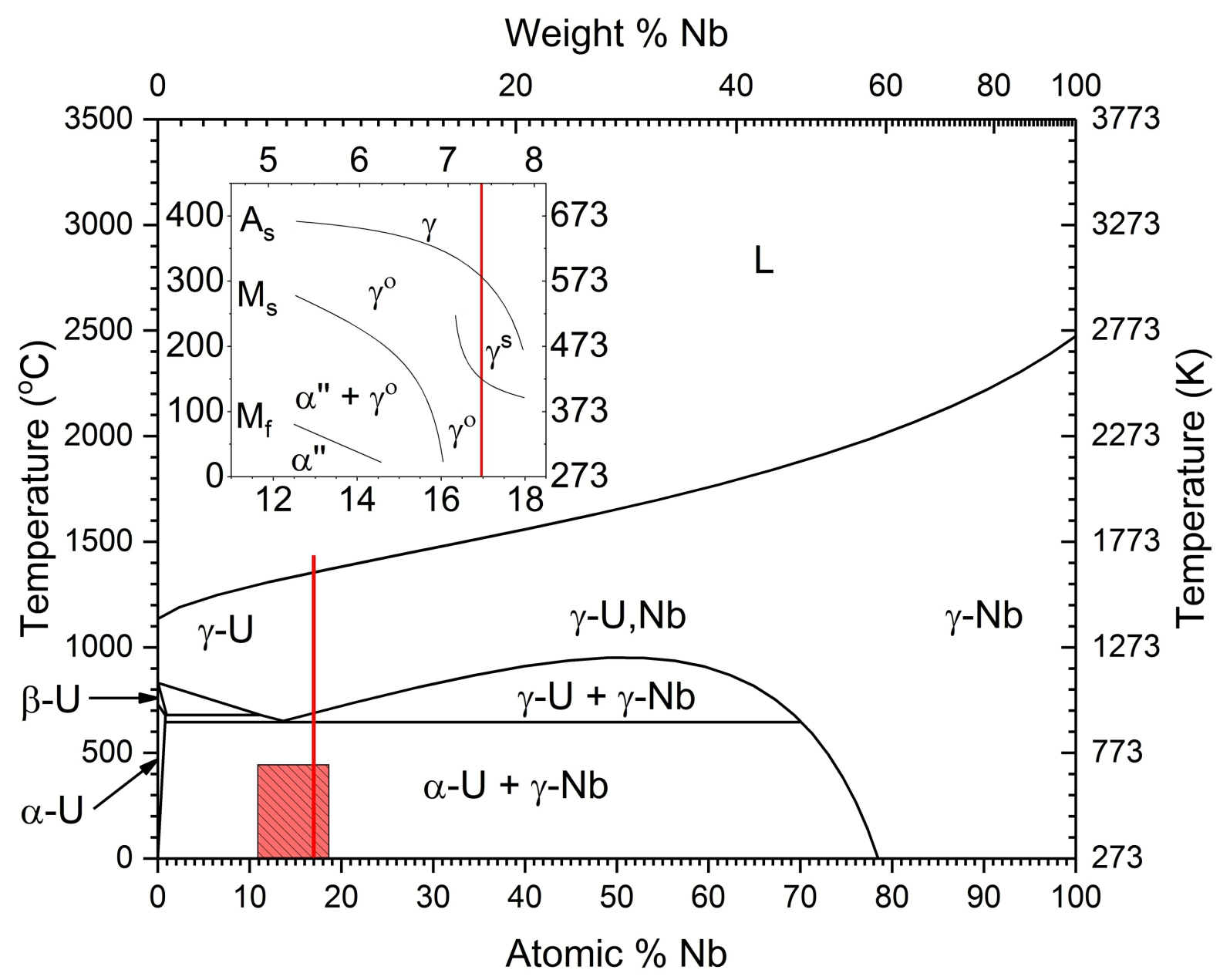}
\caption{Phase diagram of the UNb system. Inset shows the metastable states produced in the uranium-rich alloys produced by water quenching in the region denoted by the red box. Image constructed from the data of: \cite{Koike1998,Liu2008,Vandermeer1981}.} \label{fig:Phase_1}
\end{figure}

Binary phase diagrams of uranium and transition metals are rich in physics with subtleties arising between alloying species \cite{Rough1958}. The U-Nb system is unique in its lack of intermetallic compounds, Figure \ref{fig:Phase_1}. Niobium is highly soluble with uranium in the high temperature $\gamma$ phase (bcc) forming a continuous solid solution above 950$^{o}$C. Niobium is much less soluble within the $\alpha$ phase, limited to 0.5wt\%Nb \cite{Manner1999}, Figure \ref{fig:Phase_1}. Cooling U-Nb below the eutectoid at 647$^{o}$C causes phases to separate into a virtually pure uranium, $\alpha$-U, and niobium with a solubility of $\sim$40wt\%U, $\gamma$-(Nb,U) \cite{Koike1998}. The dual phase has a number of undesirable properties, such as poor corrosion resistance and mechanical properties \cite{Volz2007}.

Due to slow diffusion of transition metals in uranium \cite{Fedorov1972}, alloys with at least 7wt\%Nb are capable of producing a body centred cubic $\gamma^{s}$ phase when quenched in water from the high temperature phase, $\gamma$-U \cite{DAmato1964}. Under low temperature ageing, the $\gamma$ phase undergoes a displacement ordering transformation, shearing the $bcc$ phase, creating a body centred tetragonal $\gamma^{0}$ phase \cite{Yakel1969}. This work seeks to determine the nature of the stability of the $\gamma^{o}$ phase and whether, by accelerating the ageing process through increased temperatures, a transformation from the $\gamma^{o}$ to the $\alpha$-U $+$ $\gamma$-(Nb,U) arrangement, shown in the inset of Figure \ref{fig:Phase_1}, could be observed.

\section{Experimental}
Starting materials supplied by AWE were cast from melt with 7.5\% niobium by weight, henceforth referred to as UNb7. Carbon is the most significant impurity, 153$\pm$28 ppm, followed by oxygen, 42$\pm$4,  and nitrogen, $<$4 ppm. Following casting, a heat treatment was applied whereby the alloy was hot rolled to approximately 50\% of the original thickness. A further vacuum homogenisation treatment at 1000$^{o}$C was applied for 6 hours followed by a 850$^{o}$C cooling step and oil quench.

Ageing at 150$^{o}$C was undertaken for 1500, 3624 and 5000 hours. For comparison, this study also investigates samples of the same providence but without the high temperature ageing.

\subsection{Sample Preparation}
All samples were initially prepared for investigation via a two step procedure of mechanical and electro-polishing as outlined in Jones et al. \cite{Jones2015}.

Argon ion etching was performed at 45$^{o}$ under UHV conditions to finish the preparation process. Etching at 4\,kV with a beam current of 4.2\,$\mu$A was found to be effective in removing oxide formed whilst preserving the material underneath. Auger electron spectra confirmed that the oxide layer was completely removed.

X-ray diffraction was carried out using Cu-K$\alpha$ radiation on a Philips X'Pert Pro diffractometer. The x-ray tube was operated at 40\,kV and 30\,mA. Specular $\theta$/2$\theta$ scans were performed between 20\,$^{o}$ and 50\,$^{o}$ with a step of 0.02$^{o}$. A thin coating of CeO$_{2}$ was applied to the sample to provide calibration of the diffraction spectra.

Electron microscopy and EBSD were performed using a Zeiss EVO MA10 SEM fitted with a LaB$_{6}$ electron source, a Digiview 3 high speed camera and EDAX EBSD instrumentation. OIM$^{TM}$ software recorded and processed the data. Orientation maps of up to 1\,mm$^{2}$ were obtained with step sizes of less than 2\,$\mu$m.
$\alpha$-U, $\alpha''$, $\gamma^{o}$ and $\gamma$-U phases were used as reference phases to index against. Clean-up functions using OIM$^{TM}$ software were implemented when necessary.

TEM sections, generated from unaged and 5000 hr aged alloy samples, were extracted via FIB liftout using a Helios Nanolab 600i Dualbeam (DB). This is a two stage process utilising a high beam current to produce an initial lamella of material which is lifted out and secured to a Cu grid through ion-assisted platinum deposition before continuing thinning down to 50-100nm at progressively lower accelerating voltages. Additionally, a high sensitivity Oxford Instruments INCA X-MAX module fitted to the DB and complementary software enabled energy dispersive x-ray spectroscopy (EDS).

DB milled slices were transferred to a Philips EM430 TEM with air exposure limited to less than 10 minutes. Images were taken predominately in bright field mode using an accelerating voltage of 250\,kV with diffraction patterns taking place using a 20\,$\mu$m selected area aperture and parallel (defocused) electron beam. Diffraction patterns were mostly recorded on areas less than 10\,$\mu$m x 10\,$\mu$m, with the incident beam along prominent zone axes.

Hardness measurements were performed on polished samples of all ageing treatments using a Vickers microhardness tester fitted with a 200\,g load.

\section{Results \& Discussion} 
\subsection{Phase Identification}
\subsubsection{$\gamma$ Phases}
\begin{figure}
\centering
\includegraphics[width=\linewidth]{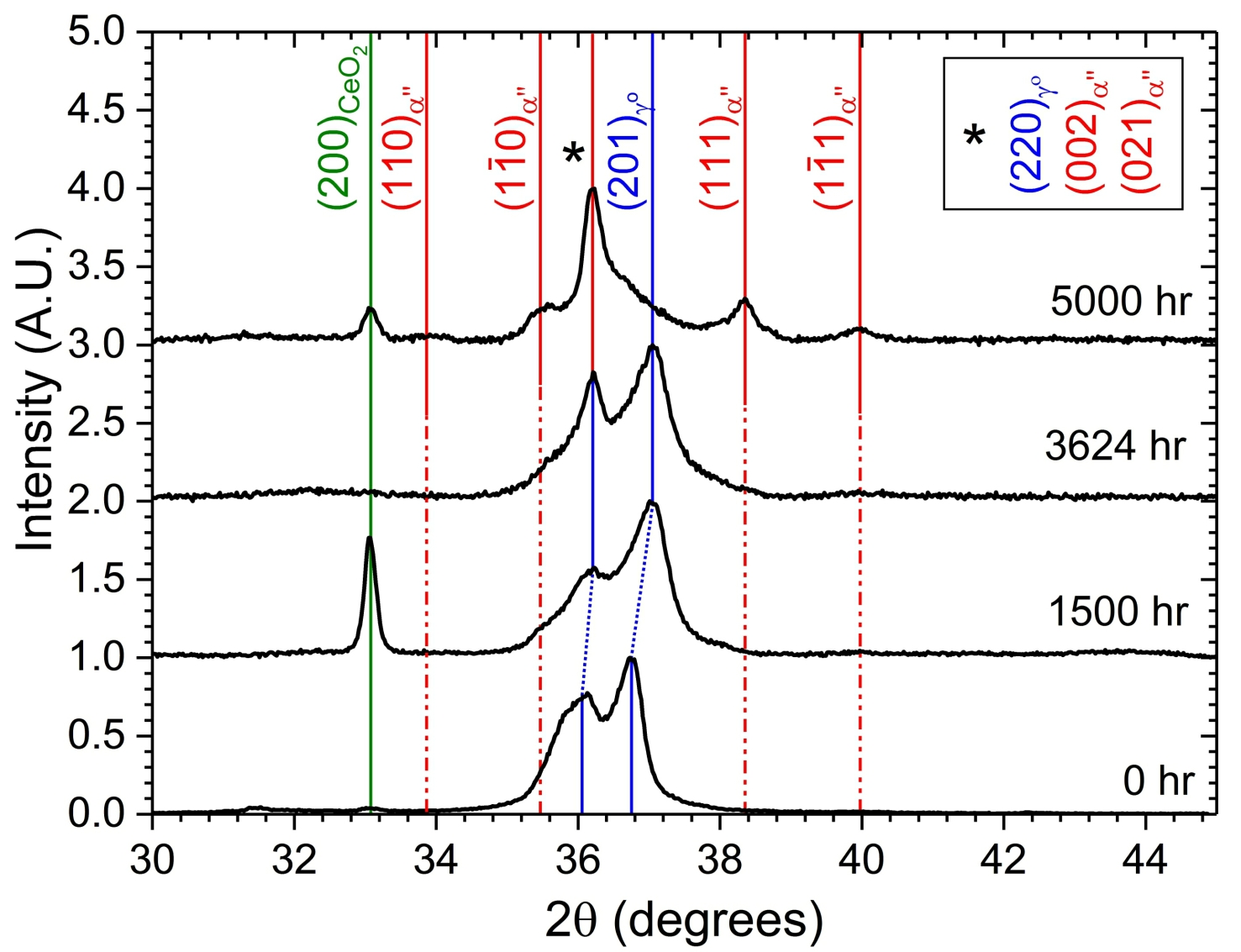}
\caption{X-ray diffraction patterns of samples aged for varying durations at 150$^{o}$C. Under ageing, the $\alpha''$ phase begins to form at the expense of the $\gamma^{o}$ phase which is also observed to shift in position and broaden.} \label{fig:XRD}
\end{figure}

XRD Bragg peaks corresponding to the $\gamma^{o}$ phase \cite{Jackson1970}, were identified in the unaged material, Figure \ref{fig:XRD}. Through ageing, stresses and strains manifested themselves, resulting in distortion of the $\gamma^{o}$ phase, producing broader peaks at greater values of 2$\theta$. No evidence was obtained for the presence of retained $\gamma$-U or $\gamma^{s}$ phases.

\begin{figure*}
\centering
\includegraphics[width=\linewidth]{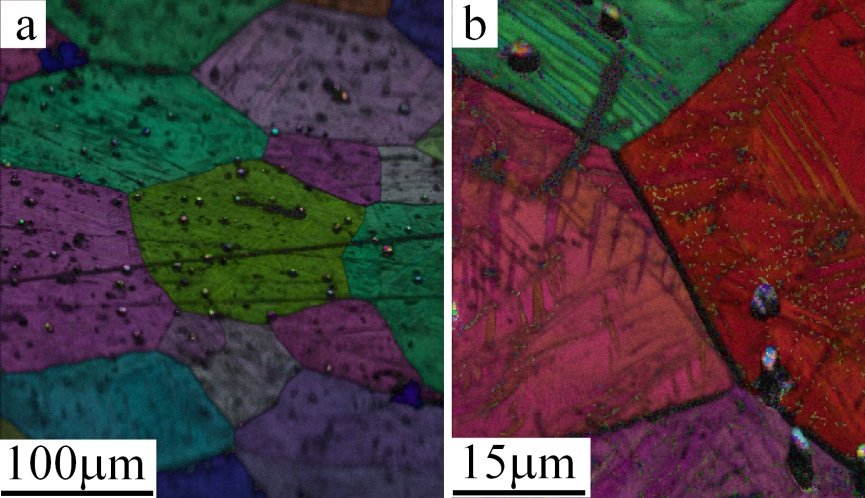}
\caption{a) Inverse pole figure map of the unaged sample showing the variety of grain orientations present. b) Inverse pole figure map of the 5000 hr aged sample highlighting the increase of intra-granular distortion.} \label{fig:EBSD}
\end{figure*}

EBSD maps, Figure \ref{fig:EBSD}, were indexed against $\gamma$-U. The higher symmetry cubic phase produced better patterns for assessing micro-structural detail than the tetragonal $\gamma^{o}$ phase which suffered from the pseudosymmetry problem. Due to the small interaction volume and a $c$/$a$ ratio very close to $1/\sqrt{2}$, indexing against the $\gamma^{o}$ phase could lead to ambiguity of crystallographic directions.

XRD identifies the phase as $\gamma^{o}$ suggesting that the EBSD patterns observed and indexed as $\gamma$ are generated by the $\gamma^{o}$ phase but the subtleties of EBSD obscure the difference as discussed above. The 1500 and 3624 hr aged samples were indexed predominately as the $\gamma$ phase with isolated regions such as twin boundaries assessed as the most likely candidates for nucleation of the $\alpha''$ phase.

\subsubsection{$\alpha$ Phases}
XRD peaks characteristic of the $\alpha''$ phase are strongly present in the 5000 hr aged sample \cite{Anagnostidis1964,Tangri1965}, Figure \ref{fig:XRD}. Peaks relating to the breaking of the \emph{bcc} symmetry start to emerge between 3624 and 5000 hrs. The transformation observed is akin to that of the athermal or a mechanically induced martensitic reaction \cite{Zhang2015}. In both these mechanisms, the $\gamma^{o}$ lattice grows in the $a$ axis and contracts in $c$ axis before undergoing a symmetry change to produce the monoclinic $\alpha''$ phase. Under the transformation, the lattice and atomic positions follow the same route as in quenching for an alloy of around 5\%wt Nb. This is observed by shifting diffraction peaks (expanding and contracting) and broadening as the population of lattice planes undergo the transition. Sharpening of peaks occurs subsequently as a large percentage of the material reaches the eventual phase, $\alpha''$, as appears to be occurring in the 5000 hr sample. Utilising the Scherrer equation \cite{Patterson1939}, the average crystallite size of the $\alpha''$ phase appears to grow with ageing as might be expected if transformed regions nucleate further $\alpha''$ phase.

EBSD of the lesser aged samples showed the $\alpha''$ phase to be dispersedly located in small subgrains less than a 1\,$\mu$m in diameter. Stresses assumed to be in the $\alpha''$ phase and the small atomic displacements which occur in the $\gamma^{o}$ to $\alpha''$ transition are believed to be responsible for poorly resolved patterns of this phase. EBSD of the 5000\,hr aged sample suggested $\gamma^{o}$ phases remained as the dominant phase, however, without a suitable material file for the as transformed $\alpha''$ phase and similar crystal structures, the $\gamma$ phase provided the better fit permitting a stronger map for the assessment of micro-structural detail.

\begin{figure}
\centering
\includegraphics[width=\linewidth]{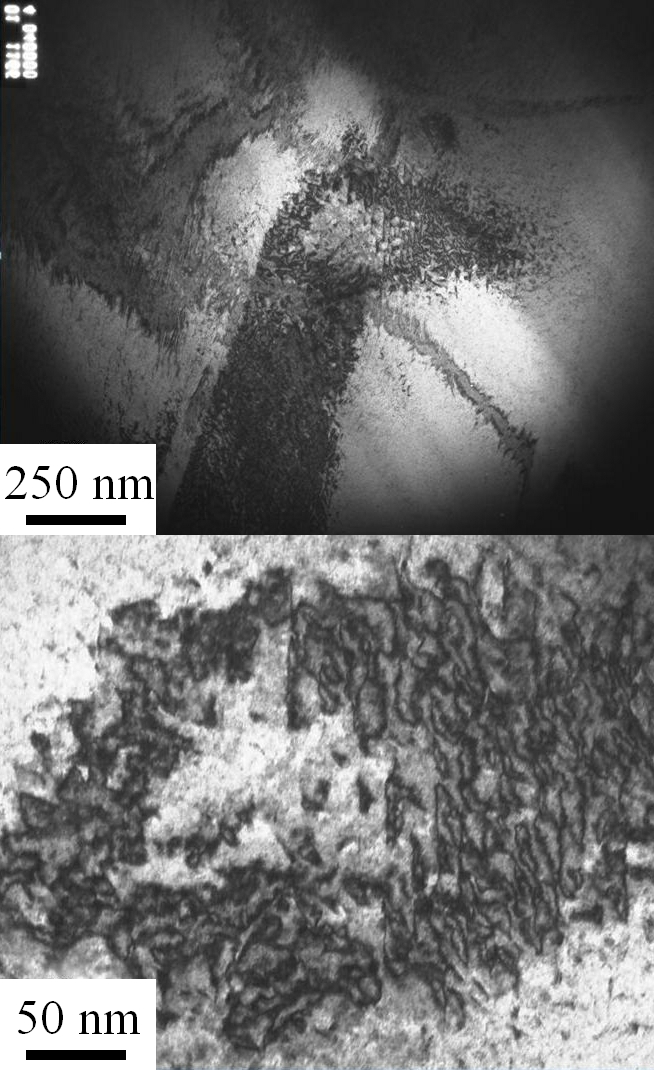}
\caption{TEM images of the 5000hr aged alloy with dark field images showing defects in crystal structure. The mottled appearance is assumed to be due to dislocation loops encompassing platelets of retained $\gamma^{o}$ in a matrix of $\alpha''$.} \label{fig:TEM_Decomp}
\end{figure}

Using high magnification TEM, the 5000\,hr aged sample was observed to possess regions with high levels of contrast when viewed in dark field, a feature usually produced by crystal defects. Figure \ref{fig:TEM_Decomp} shows angular cells bound by dark dislocation loops thought to be retained $\gamma^{o}$ phase in a matrix of predominately $\alpha''$. 

The broadening of $\gamma^{o}$ Bragg peaks observed in XRD as a function of ageing suggests diminishing volumes of coherently scattering $\gamma^{o}$ crystallites. This is illustrated in Figure \ref{fig:TEM_Decomp} where the retained $\gamma^{o}$ is very small but much more noticeable due to the high density of dislocation loops.

It is clear that the initial $\gamma^{o}$ phase has transformed into the $\alpha''$ over a period of months at a temperature of 150\,$^{o}$C. The resulting crystal structures are consistent with those produced through the athermal reaction experienced in alloys containing less niobium during manufacture \cite{Anagnostidis1964,Tangri1965}.

\subsection{Microstructural Properties}

All samples studied had very similar grain shapes and sizes, irrespective of ageing. Low temperature ageing experienced in this study did not act to anneal the sample in a way that promoted grain growth. In cross-section, grains have straight edges and are surrounded usually by 6 or 7 others. With superior hardness and electrochemical resistance relative to the metal, carbide inclusions sit proud of the polished surface.

\begin{figure}
\centering
\includegraphics[width=\linewidth]{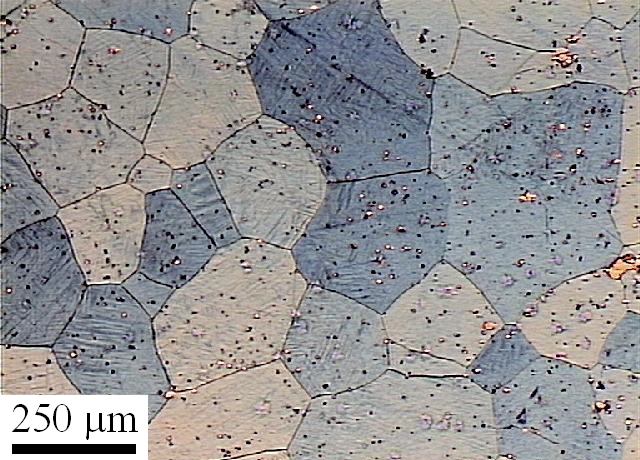}
\caption{Optical image of the 5000 hr aged sample similar to those taken through SEM images. Visible discolouration appears to be accentuated with more complicated microstructures.}\label{fig:Optical}
\end{figure}

Optical images revealed discolouration of aged polished samples exposed to air over a period of 2 hours, Figure \ref{fig:Optical}. Unaged materials were not observed to share this behaviour, in keeping with the expectation that more complex intergranular structures lead to oxidation rate anisotropy \cite{Cathcart1972}.

\begin{figure}
\centering
\includegraphics[width=\linewidth]{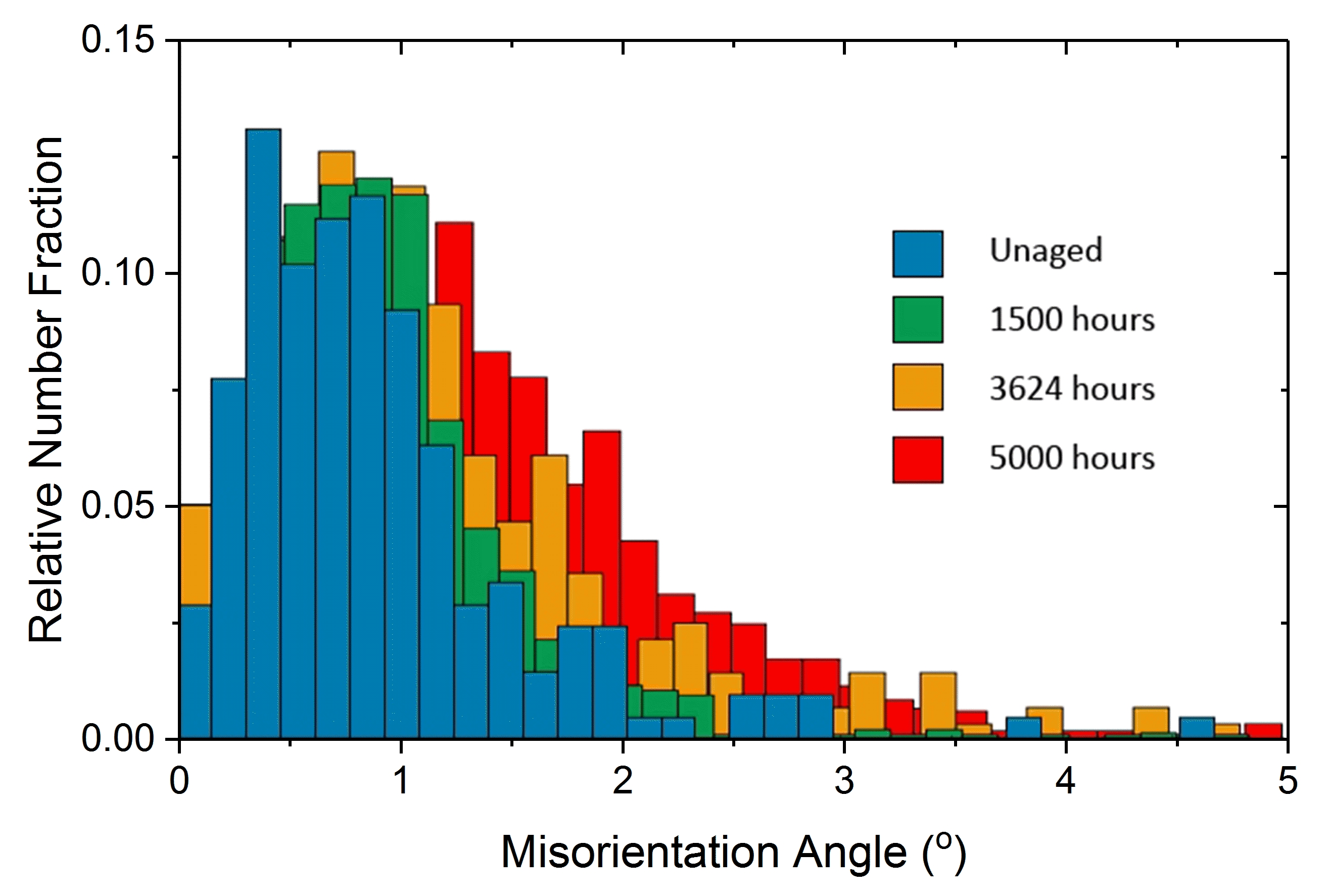}
\caption{Grain average misorientation plot focusing on the kinking of grains.}\label{fig:GAM}
\end{figure}

Intra-granular characteristics were observed to evolve with ageing. Lineations, initially assumed to be stress induced $\gamma^{o}$ kinking became more numerous and larger with greater ageing, Figure \ref{fig:Optical}. Grain average misorientation charts, Figure \ref{fig:GAM}, were generated from regions encompassing numerous grains, such as those shown in Figure \ref{fig:EBSD}, to quantify the discrepancy in orientations of neighbouring points. Distributions were normalised in the 0-5$^{o}$ range to focus on kinking rather than twinning \cite{Teeg1961}, and were observed to behave as Poisson-like with the mean increasing from 0.8 to 1.1$^{o}$ with ageing. This confirms that ageing had the effect of producing kink defects, disrupting intra-granular arrangements \cite{Teeg1961,Cahn1953}.

\begin{figure}
\centering
\includegraphics[width=\linewidth]{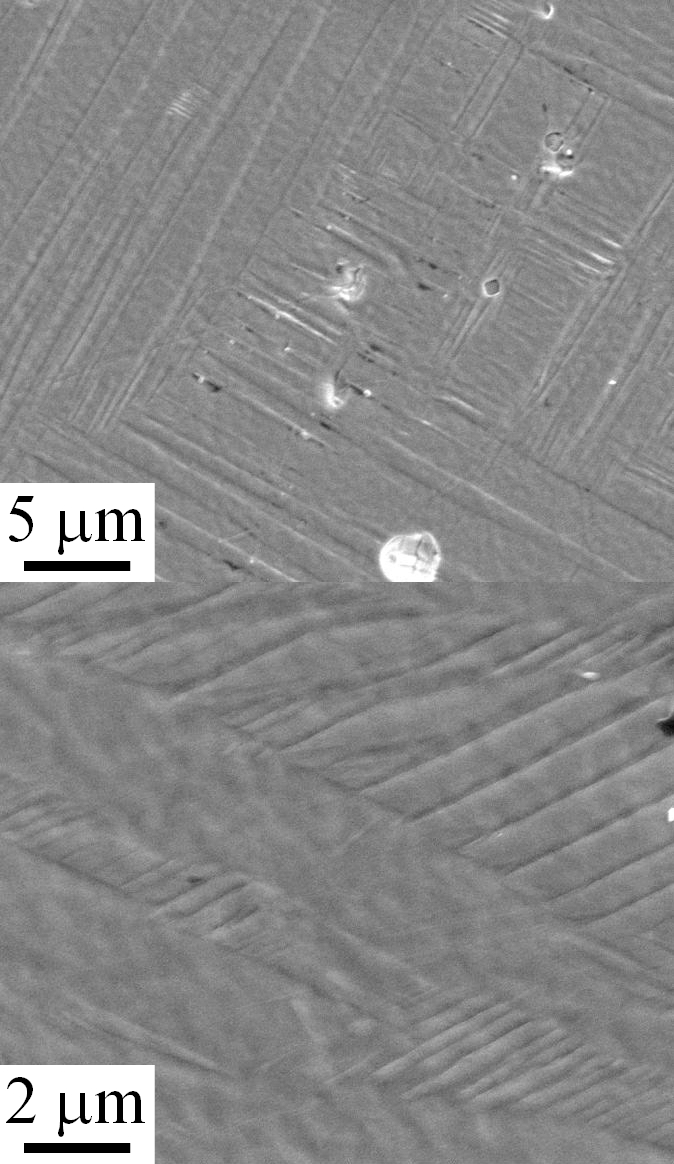}
\caption{SEM images of the surface of the 5000 hr aged sample. Plates visible are very similar to those expected by the martensite reaction. Preferential etching of the chemically weaker twin/martensite variant boundaries emphasises intra-granular structures during the polishing process.}\label{fig:SEM}
\end{figure}

\begin{figure}
\centering
\includegraphics[width=\linewidth]{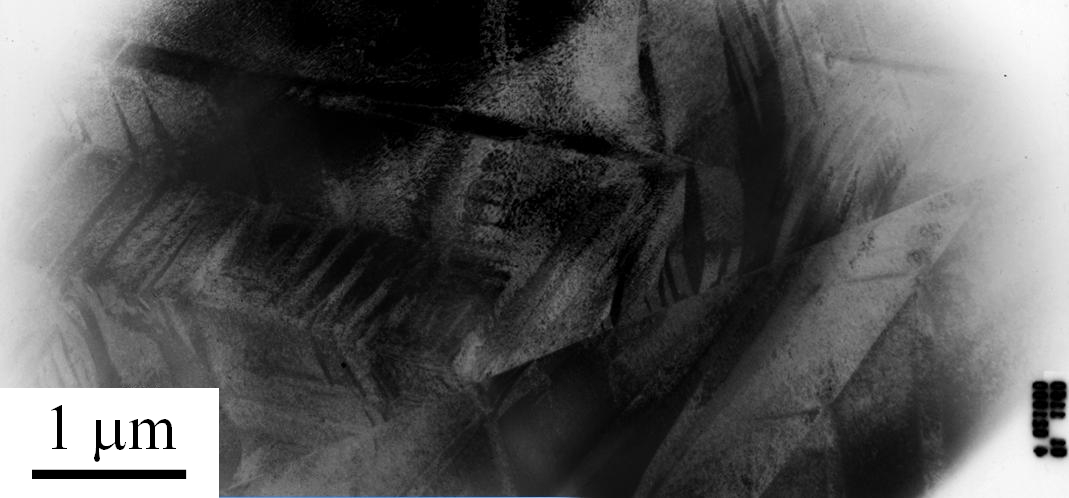}
\caption{Bladed twin-type structures visible in TEM images of the 5000hrs aged sample.} \label{fig:TEM_Blade}
\end{figure}

Martensitic herringbone twinning structures such as those observed in Figures \ref{fig:SEM} and \ref{fig:TEM_Blade} are known to manifest themselves in the $\alpha''$ phase \cite{Hatt1966,Field2001,Field2013,Zuev2013}, and became more prevalent in the SEM and optical images of the 5000\,hr aged sample. Twinning and slip defects are a regular occurrence in $\alpha$-U, resulting from the anisotropic thermal expansion of $\alpha$ grains \cite{Cahn1953,Zuev2013}. A periodicity of around 100 nm was observed showing a close resemblance to those seen previously \cite{Field2001,Zuev2013}. 

Under TEM examination, the 5000\,hr aged sample showed greater degrees of micro and nanostructural distortion than the unaged case. Fine nanoscale twins, assumed to be alternating orientations of $\alpha''$ phase, were observed to be discontinuous across major twins producing characteristic herringbone twinning and blade-like structures, Figure \ref{fig:TEM_Blade}. Branching and coarsening of these martensitic twins could also be observed as they approached twin boundaries which may point to regions of retained `austenitic' $\gamma^{o}$ \cite{Kohn1992}. Extinction contours were displaced over twin-type structures suggesting that rotational lattice displacements of around 1-10$^{o}$, in keeping with kinking observed using EBSD, were present on this scale, possibly due to new twin formation \cite{Crocker1976}. Intensive twinning and `mottling' at far smaller scales become particularly prevalent with ageing, Figure \ref{fig:TEM_Decomp}.

The spatial occurrence of the $\alpha''$ phase has not been definitely located in this study. Using EBSD, it is possible to discern small, sparsely distributed streaks of $\alpha''$ phase. The $\alpha''$ appears to have formed in the thinnest internal kinks, assumed to be the ordered $\gamma^{o}$ phase, Figure \ref{fig:EBSD}. It is likely that the $\alpha''$ phase is nucleating at the boundaries of the thinnest twins where stress is assumed at a maximum.

Electron microscopy shows evidence of a build up of stresses in the material, in keeping with XRD observations. XRD data strongly resembles a recent work by Zhang \cite{Zhang2015}, in which the UNb7 alloy was loaded with tensile strain and a phase change occurred due to a stress induced martensitic reaction. Martensitic reactions can be stimulated by a variety of factors including temperature, stress and magnetic fields in ferromagnets \cite{Lobodyuk2005}. The data presented in this work suggest that modest temperatures (150 $^{o}$C) encourages a stress-induced isothermal martensitic transition in a way similar to external mechanical deformation. Since the UNb7 martensitic transformation temperature has been measured to exist well below 0$^{o}$C, Figure \ref{fig:Phase_1}, it is unlikely that an athermal martensitic transformation to the $\alpha''$ has been induced at any point in the material's lifetime.

TEM observations agree with the conjecture that the martensitic reaction is responsible for the phase transformation producing structures such as the characteristic twinning and resultant bladed structures such as those in Figure \ref{fig:TEM_Blade}. Martensitic twinning is obvious in these regions as well as potential evidence of chemical segregation at boundaries. A small amount of diffusion might be responsible for initiating the transformation as in the case of a massive reaction \cite{Burke1965}. Additionally, superlattice-esque reflections observed through TEM diffraction, Figure \ref{fig:TEM_Banding}, suggest crystallographic relations exist between phases, though these have not been extensively studied and will form the basis of future work. Possible mechanisms for the appearance of superlattice formation include $\gamma$ $\rightarrow$ $\gamma^{o}$ or $\gamma^{o}$ $\rightarrow$ $\alpha''$ martensitic transformations, or successive twin variants of the $\alpha''$ phase.

\subsection{Chemical Composition}
EDX line scans were applied to the vicinity of grain boundaries and intra-granular features such as kinks and twins in SEM and STEM modes. No strong evidence was found via this method to suggest niobium clustering at crystallographic defects, as ageing progresses. Due to similar electron densities and crystallographic structures between the $\gamma^{o}$ and $\alpha''$ phases, no contrast can be observed via SEM backscatter detection.

\begin{figure*}
\centering
\includegraphics[width=\linewidth]{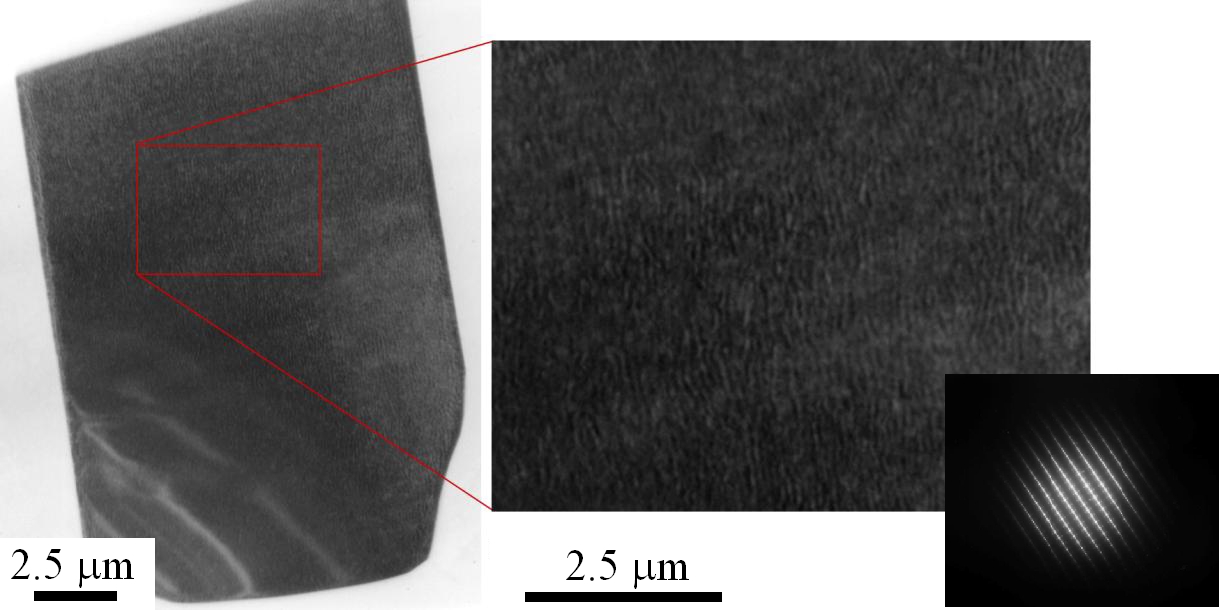}  
\caption{TEM section of the unaged UNb7 alloy taken from a single crystal grain. A large area complementary diffraction pattern alternating orientations of $\alpha''$ twins or $\gamma^{o}$/$\alpha''$ are giving the impression of a superlattice structure.} \label{fig:TEM_Banding}
\end{figure*}

Bright-field TEM images, such as those in Figures \ref{fig:TEM_Decomp} \& \ref{fig:TEM_Banding}, displayed bands alternating in contrast. Bright field contrast is often ascribed to differences in electron density and so could suggest chemical segregation with light bands indicative of Nb-enriched and dark bands of Nb-deficient regions. However, relative rates of channelling of electrons between phases and orientations also affects contrast in bright field mode and is thought to be responsible in this case; twinning and martensitic structures are known to be prevalent on the same scale from dark field images. Additionally, straight lines suggest that the bands are following specific lattice directions and therefore crystallographic features. Speckling within crystal twins such as those featured in Figure \ref{fig:TEM_Blade} may present some evidence for solute segregation on the nanoscale.

Early works evaluating decomposition routes has shown that similarities exist between a large proportion of the uranium alloy systems \cite{Cathcart1972,Chiswik1958,White1955,Holden1953,Vandermeer1973}. Two time-temperature-transition C-curves are often observed with an upper transition above 300$^{o}$C clearly driven by diffusion and a lower curve existing below 300$^{o}$C. The mechanism behind this lower curve has been a subject of debate recently with a number of mechanisms having been proposed as responsible for the evolution of UNb alloys at low temperatures \cite{Clarke2009}.

Were the lower reaction a diffusion controlled transition, regions of $\gamma^{s}$ and/or $\gamma-U$ phase would be produced through ageing as regions of the material diverge in composition. Diffusional mechanisms, particularly eutectoid decomposition, leads to separation of constituents obvious from bands readily detectable by a range of techniques. Very little evidence in this study has been found to suggest that clustering of solute is taking place. Furthermore, by application of the lever rule to this system the finishing point of a diffusional reaction can be approximated. It would be expected that, provided no inter-metallics are present in the UNb system and for which there has been no evidence \cite{Koike1998}, the finishing ratio would be roughly 88\%wt $\alpha$-U : 12 \%wt $\gamma$-(Nb,U). This is roughly the ratio of the $\alpha''$ to $\gamma^{o}$ phases observed in the 5000 hr aged case through XRD, however the phases do not correspond to those expected by diffusional means, suggesting that another reaction is responsible.

The coherent spinodal has been calculated to pass close to UNb7 at room temperature suggesting that nucleation and growth, and spinodal decomposition mechanisms are likely to be convoluted, particularly in an imperfect system \cite{Clarke2009}. In reality, the idea of a distinct separation of spinodal decomposition and nucleation and growth relies on mean field theory which does not truly hold for metallic alloys \cite{Cahn1996}. It is true that in certain parts of the phase diagram, nucleation and growth and spinodal decomposition may be more prevalent, however experimental work has shown an overlap between mechanisms in alloy systems \cite{Cahn1996}. Moreover, the validity of a chemical spinodal for metastable and martensitic phases is also questionable given the stressed nature of the alloy post quench and numerous crystallographic features and defects.

Vandermeer's study of a more dilute UNbZr alloy suggested that spinodal decomposition took place before a stress-induced martensitic transformation changed the phase of the material \cite{Vandermeer1973}. This second step is assumed to be the same effect seen in this study. Potentially, since the composition studied in this paper lies so close to the martensitic transition both in terms of temperature and composition, the isothermal martensitic transition occurs without a necessary shedding of niobium from the matrix.

\subsection{Mechanical Properties}
\begin{figure}
\centering
\includegraphics[width=\linewidth]{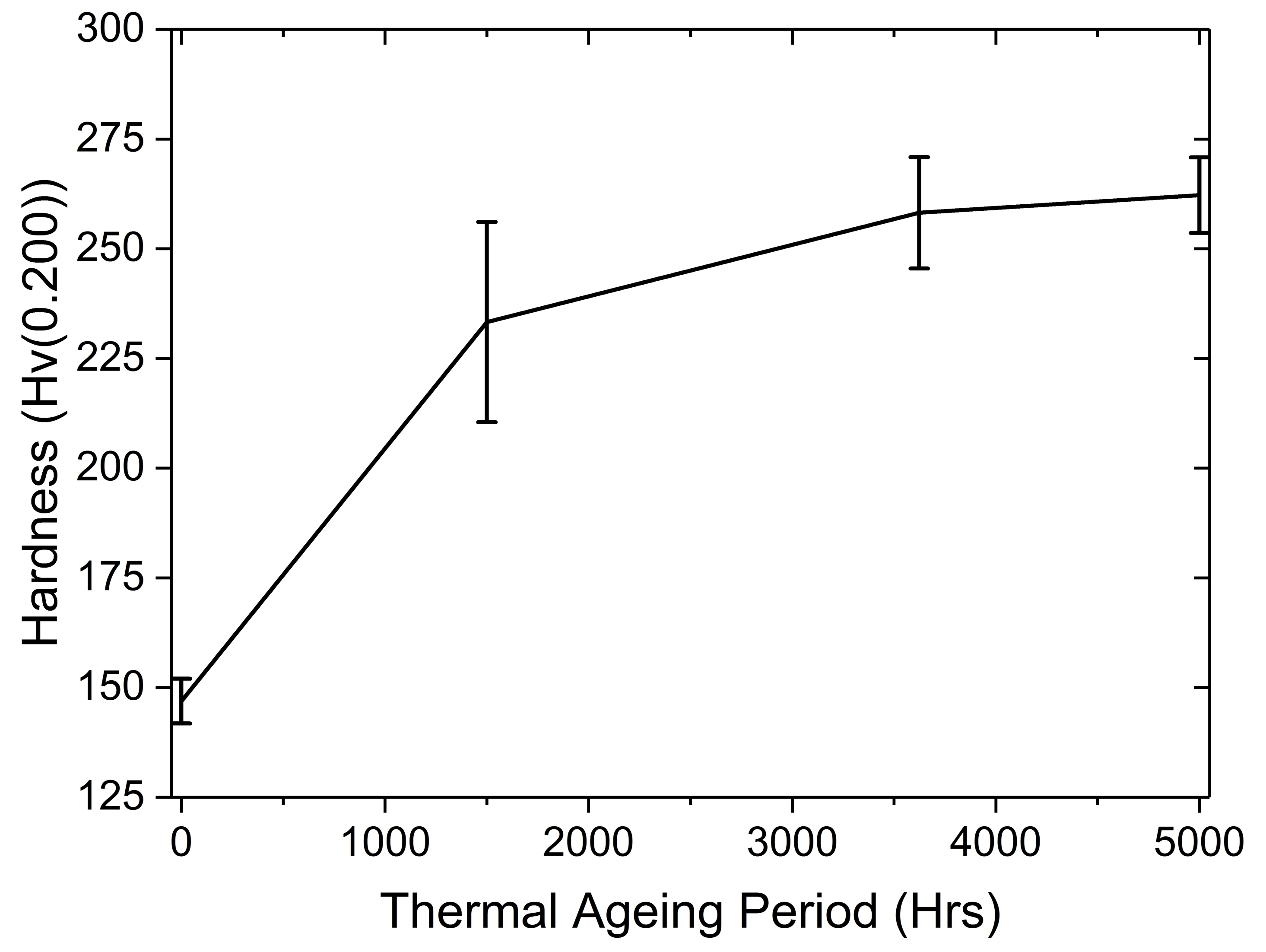}
\caption{Evolution of hardness over ageing period as determined by the Vickers method.} \label{fig:Vickers}
\end{figure}

Vickers hardness testing of UNb7 observed an increase with ageing before plateauing at greater ageing durations, Figure \ref{fig:Vickers}. Since, the unaged sample had undergone 6 years worth of ambient ageing, it is important to compare the result with that expected post-manufacture.  The value of 146.9$\pm$5.1 Hv as observed in this study is reasonable and consistent with other studies \cite{Wheeler2009,Volz2007}, suggesting that any ambient ageing that occurred in this time has had little effect on the hardness.

\begin{figure}
\centering
\includegraphics[width=\linewidth]{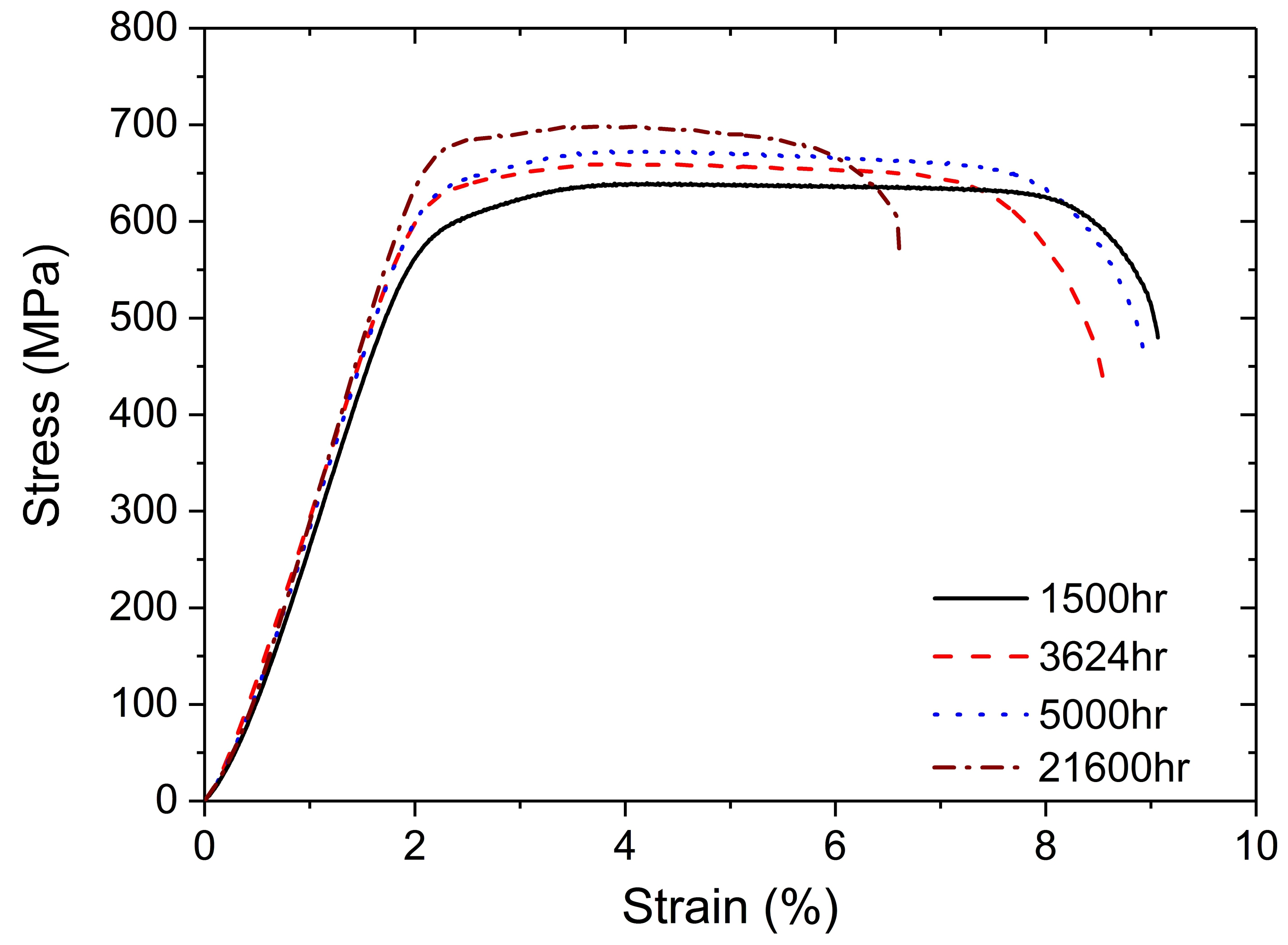}
\caption{Stress-strain curves of tensile test specimens that have been aged to the three ageing conditions plus an additional period of 21600 hours.} \label{fig:Stress-Strain}
\end{figure}

With ageing, UNb7's hardness is greater than unaged UNb6 and UNb5, becoming comparable to that of unalloyed uranium. The hardness of UNb7 exceeds unaged martensitic UNb5 between 0 and 1500 hours, consistent with the expectation that dislocations are forming due to shearing of the lattice under the martensitic reaction and the building up of twins thereby inhibiting the slipping of grains and increasing hardness. Both transformations are well known to produce a hardening effect \cite{Nishiyama1978}.

Stress-strain curves from ambient tensile testing shows the yield strength increasing with ageing, consistent with hardness testing. Young's modulus remains relatively constant in all cases suggesting no change in bulk phase but the evolution of a defect structure inhibiting dislocation motion contributing to an increased ultimate strength and hardness. 

The 21600 hr sample weakly exhibited double yielding. Although the material is starting to share characteristics with the martensitic $\alpha''$ phase, strong evidence for detwinning should not be expected to present itself in stress-strain curves. Twinning observed in this study has been confirmed with the use of TEM. Figure \ref{fig:EBSD}b shows the increase of twin propensity. In comparison, the twinning occurring in the UNb5 system is on much larger length scales and therefore likely to be a much more significant effect on macro-mechanical properties.

\section{Conclusion}
This work has established that the microstructure of this alloy is both heterogeneous and convoluted, possessing several deformation routes, possibly driven by different decomposition regimes. Through ageing, the microstructure becomes more complex developing a greater density of twins and kinks across a wide length scale.

Little evidence was found for appreciable solute clustering. Possible evidence for clustering can be found from selected regions such as twin boundaries and $\gamma^{o}$-$\alpha''$ phase boundaries as observed by TEM. However, the phase transition has been dominated by an isothermal martensitic transformation. It should be investigated further whether minute diffusion of atoms is sufficient to provide enough instantaneous energy for the material to undergo the transition.

Ageing has had a profound effect on the mechanical properties of this system as would be expected. Hardness and ductility were measured and assessed to have followed a path that is expected of stress-induced isothermal martensitic transformation.

\bibliographystyle{unsrt}
\bibliography{U7Paper}

\begin{thebibliography}{10}

\bibitem{Snelgrove1997}
J.L. Snelgrove, G.L. Hofman, M.K. Meyer, C.L. Trybus, and T.C. Wiencek.
\newblock {Development of very-high-density low-enriched-uranium fuels}.
\newblock {\em Nucl. Eng. Des.}, 178(August):119--126, 1997.

\bibitem{Meyer2002}
M.~K. Meyer, G.~L. Hofman, S.~L. Hayes, C.~R. Clark, T.~C. Wiencek, J.~L.
  Snelgrove, R.~V. Strain, and K.~H. Kim.
\newblock {Low-temperature irradiation behavior of uranium-molybdenum alloy
  dispersion fuel}.
\newblock {\em J. Nucl. Mater.}, 304(2-3):221--236, 2002.

\bibitem{Savchenko2010}
A.~Savchenko, A.~Vatulin, I.~Konovalov, A.~Morozov, V.~Sorokin, and
  S.~Maranchak.
\newblock {Fuel of novel generation for PWR and as alternative to MOX fuel}.
\newblock {\em Energy Convers. Manag.}, 51(9):1826--1833, 2010.

\bibitem{Sinha2009}
V.~P. Sinha, G.~J. Prasad, P.~V. Hegde, R.~Keswani, C.~B. Basak, S.~Pal, and
  G.~P. Mishra.
\newblock {Development, preparation and characterization of uranium molybdenum
  alloys for dispersion fuel application}.
\newblock {\em J. Alloys Compd.}, 473(1-2):238--244, 2009.

\bibitem{VandenBerghe2008a}
S.~{Van den Berghe}, W.~{Van Renterghem}, and a.~Leenaers.
\newblock {Transmission electron microscopy investigation of irradiated
  U–7wt{\%}Mo dispersion fuel}.
\newblock {\em J. Nucl. Mater.}, 375(3):340--346, 2008.

\bibitem{Clarke2015}
A.J. Clarke, K.D. Clarke, R.J. McCabe, C.T. Necker, P.A. Papin, R.D. Field,
  A.M. Kelly, T.J. Tucker, R.T. Forsyth, P.O. Dickerson, J.C. Foley,
  H.~Swenson, R.M. Aikin, and D.E. Dombrowski.
\newblock {Microstructural evolution of a uranium-10 wt {\%} molybdenum alloy
  for nuclear reactor fuels}.
\newblock {\em J. Nucl. Mater.}, 465:784--792, 2015.

\bibitem{Wheeler2009}
D.~W. Wheeler and S.~T. Morris.
\newblock {Micro-mechanical characterisation of uranium}.
\newblock {\em J. Nucl. Mater.}, 385(1):122--125, 2009.

\bibitem{Jackson1970}
R.~J. Jackson.
\newblock {Reversible martensitic transformations between transition phases of
  uranium-base niobium alloys}.
\newblock Technical report, The DOW Chemical Company, Rocky Flats Division,
  Golden, Colorado, 1970.

\bibitem{Manner1999}
W.L. Manner, J.A. Lloyd, R.J. {Hanrahan Jr.}, and M.T. Paffett.
\newblock {An examination of the initial oxidation of a uranium-base alloy
  (U–14.1 at.{\%} Nb) by O2 and D2O using surface-sensitive techniques}.
\newblock {\em Appl. Surf. Sci.}, 150(1–4):73--88, 1999.

\bibitem{Rough1958}
F.A. Rough and A.A. Bauer.
\newblock {Constitution of uranium and thorium alloys}.
\newblock Technical report, Battelle Memorial Institute, Columbus, Ohio, 1958.

\bibitem{Koike1998}
J.~Koike, M.E. Kassner, R.E. Tate, and R.S. Rosen.
\newblock {The Nb-U (niobium-uranium) system}.
\newblock {\em J. Phase Equilibria}, 19(3):253--260, 1998.

\bibitem{Volz2007}
H.~M. Volz, R.~E. Hackenberg, A.~M. Kelly, W.~L. Hults, A.~C. Lawson, R.~D.
  Field, D.~F. Teter, and D.~J. Thoma.
\newblock {X-ray diffraction analyses of aged U-Nb alloys}.
\newblock {\em J. Alloys Compd.}, 444-445(SPEC. ISS.):217--225, 2007.

\bibitem{Fedorov1972}
G.B. Fedorov, E.A. Smirnov, and V.N. Gusev.
\newblock {Diffusional and Thermodynamic Properties of the gamma-Phase of the
  System Uranium-Niobium}.
\newblock {\em At. Energiya}, 32(1):11--14, 1972.

\bibitem{DAmato1964}
C.~D'Amato, F~S Saraceno, and T~B Wilson.
\newblock {Phase transformations and equilibrium structures in uranium-rich
  niobium alloys}.
\newblock {\em J. Nucl. Mater.}, 12(3):291--304, 1964.

\bibitem{Yakel1969}
H~Yakel.
\newblock {Crystal Structures of Transition Phases Formed in
  U-16.6at{\%}Nb-5.64at{\%}Zr Alloys}.
\newblock {\em J. Nucl. Mater.}, 33:286--295, 1969.

\bibitem{Liu2008}
X.~J. Liu, Z.~S. Li, J.~Wang, and C.~P. Wang.
\newblock {Thermodynamic modeling of the U-Mn and U-Nb systems}.
\newblock {\em J. Nucl. Mater.}, 380(1-3):99--104, 2008.

\bibitem{Vandermeer1981}
R~A Vandermeer, J~C Ogle, and W~G Northcutt.
\newblock {A Phenomenological Study of the Shape Memory Effect in
  Polycrystalline Uranium-Niobium Alloys}.
\newblock {\em Metall. Trans.}, 12(May):733--741, 1981.

\bibitem{Jones2015}
C.P. Jones, T.B. Scott, and J.R. Petherbridge.
\newblock {Structural deformation of metallic uranium surrounding hydride
  growth sites}.
\newblock {\em Corros. Sci.}, 96:144--151, 2015.

\bibitem{Anagnostidis1964}
M.~Anagnostidis, M.~Colombie, and H.~Monti.
\newblock {Phases Metastables Dans Les Alliages Uranium-Niobium}.
\newblock {\em J. Nucl. Mater.}, 1:67, 1964.

\bibitem{Tangri1965}
K~Tangri, D~K Chaudhuri, and C~N Rao.
\newblock {Metastable phases in uranium alloys with high solute solubility in
  the BCC gamma phase. Part I — the system U-Nb}.
\newblock {\em J. Nucl. Mater.}, 15(4):288--297, 1965.

\bibitem{Zhang2015}
Yanzhi Zhang, Xiaolin Wang, Qinying Xu, and Yufei Li.
\newblock {X-ray diffraction study of low temperature aging in
  U–5.8wt.{\%}Nb}.
\newblock {\em J. Nucl. Mater.}, 456:41--45, 2015.

\bibitem{Patterson1939}
A.~L. Patterson.
\newblock {The scherrer formula for X-ray particle size determination}.
\newblock {\em Phys. Rev.}, 56(10):978--982, 1939.

\bibitem{Cathcart1972}
J.V. Cathcart and G~F Petersen.
\newblock {The Low-Temperature Oxidation of U-Nb and U-Nb-Zr Alloys}.
\newblock {\em J. Nucl. Mater.}, 43:86--92, 1972.

\bibitem{Teeg1961}
R~O Teeg and R~E Ogilvie.
\newblock {Effect of orientation and temperature on the modes of deformation of
  uranium}.
\newblock {\em J. Nucl. Mater.}, 3(1):81--88, 1961.

\bibitem{Cahn1953}
R.W. Cahn.
\newblock {Plastic Deformation of Alpha-Uranium; Twinning and Slip}.
\newblock {\em Acta Metall.}, 1(1):49--70, 1953.

\bibitem{Hatt1966}
B.A Hatt.
\newblock {The orientation relationship between the gamma and alpha structures
  in uranium-zirconium alloys}.
\newblock {\em J. Nucl. Mater.}, 19(2):133--141, 1966.

\bibitem{Field2001}
R.~D. Field, D.~J. Thoma, P.~S. Dunn, D.~W. Brown, and C.~M. Cady.
\newblock {Martensitic structures and deformation twinning in the U–Nb
  shape-memory alloys}.
\newblock {\em Philos. Mag. A}, 81(7):1691--1724, 2001.

\bibitem{Field2013}
R.~D. Field and D.~J. Thoma.
\newblock {Crystallographic and kinetic origins of acicular and banded
  microstructures in U-Nb alloys}.
\newblock {\em J. Nucl. Mater.}, 436(1-3):105--117, 2013.

\bibitem{Zuev2013}
Y~N Zuev, V~V Sagaradze, N~L Pecherkina, I~I Kabanova, I~L Svyatov, S~V
  Bondarchuk, and D~V Belyaev.
\newblock {Phase and structural transformations in uranium and uranium-niobium
  alloy upon severe deformation and heat treatments}.
\newblock {\em Phys. Met. Met.}, 114(13):1123--1154, 2013.

\bibitem{Kohn1992}
Robert~V. Kohn and Stefan M{\"{u}}ller.
\newblock {Branching of twins near an austenite—twinned-martensite
  interface}.
\newblock {\em Philos. Mag. A}, 66(5):697--715, 1992.

\bibitem{Crocker1976}
A~G Crocker and J~S Abell.
\newblock {The crystallography of deformation kinking}.
\newblock {\em Philos. Mag.}, 33(2):305--316, 1976.

\bibitem{Lobodyuk2005}
V.A. Lobodyuk and E.I. Estrin.
\newblock {Isothermal martensitic transformations}.
\newblock {\em Physics-uspekhi}, 48(7):713, 2005.

\bibitem{Burke1965}
J.~Burke.
\newblock {\em {The Kinetics of Phase Transformations in Metals}}.
\newblock Pergamon Press Ltd., 1965.

\bibitem{Chiswik1958}
H~H Chiswik, A~E Dwight, L~T Lloyd, M~V Nevitt, and S~T Zegler.
\newblock {Advances in the Physical Metallurgy of Uranium and its Alloys}.
\newblock Technical report, Second United Nations International Conference on
  the Peaceful Uses of Atomic Energy, 1958.

\bibitem{White1955}
D.W. White.
\newblock {Transformation Kinetics in Uranium-Chromium Alloys}.
\newblock {\em JOM}, 7(11):1221--1228, 1955.

\bibitem{Holden1953}
A.N. Holden.
\newblock {The Isothermal Transformation of Metastable Beta-Uranium Single
  Crystals}.
\newblock {\em Acta Metall.}, 1:617--623, 1953.

\bibitem{Vandermeer1973}
R~A Vandermeer.
\newblock {Recent Observations of Phase Transformations in a U-Nb-Zr Alloy}.
\newblock Technical report, 1973.

\bibitem{Clarke2009}
A.J. Clarke, R.D. Field, R.E. Hackenberg, D.J. Thoma, D.W. Brown, D.F. Teter,
  M.K. Miller, K.F. Russell, D.V. Edmonds, and G.~Beverini.
\newblock {Low temperature age hardening in U–13at.{\%} Nb: An assessment of
  chemical redistribution mechanisms}.
\newblock {\em J. Nucl. Mater.}, 393(2):282--291, 2009.

\bibitem{Cahn1996}
R.W. Cahn, P.~Haasen, and E.J. Kramer.
\newblock {Structure and Properties of Nonferrous Alloys}.
\newblock In {\em Mater. Sci. Technol. A Compr. Treat.} 1996.

\bibitem{Nishiyama1978}
Z~Nishiyama.
\newblock {\em {Martensitic Transformation}}.
\newblock Academic Press, New York, London, 1978.

\end{thebibliography}

\end{document}